\documentclass[aps,prl,twocolumn,superscriptaddress]{revtex4-1}
\usepackage{graphicx}
\usepackage{dcolumn}
\usepackage{bm}
\usepackage{color}
\usepackage{ulem}
\usepackage[usenames,dvipsnames,svgnames,table]{xcolor}

\newcommand{\pstar}{$p^{\star}$}
\newcommand{\psdw}{$p_{\rm SDW}$}

\newcommand{\kzero}{$\kappa_0/T$}

\newcommand{\nH}{$n_{\rm H}$}

\newcommand{\Tc}{$T_{\rm c}$}
\newcommand{\Tstar}{$T^{\star}$}
\newcommand{\Trho}{$T_{\rho}$}
\newcommand{\Tmin}{$T_{\rm min}$}
\newcommand{\Tsdw}{$T_{\rm SDW}$}

\newcommand{\Hc}{$H_{\rm c2}$}
\newcommand{\Hsdw}{$H_{\rm SDW}$}
\newcommand{\Hvs}{$H_{\rm vs}$}

\begin{document}



\title{Link between magnetism and resistivity upturn in cuprates:\\
a thermal conductivity study of La$_{2-x}$Sr$_x$CuO$_4$}

\author{P.~Bourgeois-Hope}
\affiliation{Institut Quantique, D\'epartement de physique \& RQMP, Universit\'e de Sherbrooke, Sherbrooke, Qu\'ebec, Canada J1K 2R1}

\author{S.~Y.~Li}
\altaffiliation{Present address: State Key Laboratory of Surface Physics, Department of Physics, and Laboratory of Advanced Materials, Fudan University, Shanghai 200433, China}
\affiliation{Institut Quantique, D\'epartement de physique \& RQMP, Universit\'e de Sherbrooke, Sherbrooke, Qu\'ebec, Canada J1K 2R1}

\author{F.~Lalibert\'e}
\affiliation{Institut Quantique, D\'epartement de physique \& RQMP, Universit\'e de Sherbrooke, Sherbrooke, Qu\'ebec, Canada J1K 2R1}

\author{S.~Badoux}
\affiliation{Institut Quantique, D\'epartement de physique \& RQMP, Universit\'e de Sherbrooke, Sherbrooke, Qu\'ebec, Canada J1K 2R1}

\author{S.~M.~Hayden}
\affiliation{H. H. Wills Physics Laboratory, University of Bristol, Bristol BS8 1TL, United Kingdom}

\author{N.~Momono}
\affiliation{Department of Applied Sciences, Muroran Institute of Technology, Muroran 050-8585, Japan}

\author{T.~Kurosawa}
\affiliation{Department of Physics, Hokkaido University, Sapporo 060-0810, Japan}

\author{K.~Yamada}
\affiliation{Institute for Materials Research, Tohoku University, Sendai 980-8577, Japan}

\author{H.~Takagi}
\affiliation{Department of Advanced Materials, University of Tokyo, Kashiwa 277-8561, Japan}

\author{Nicolas~Doiron-Leyraud}
\email[]{nicolas.doiron-leyraud@usherbrooke.ca}
\affiliation{Institut Quantique, D\'epartement de physique \& RQMP, Universit\'e de Sherbrooke, Sherbrooke, Qu\'ebec, Canada J1K 2R1}

\author{Louis~Taillefer}
\email[]{louis.taillefer@usherbrooke.ca}
\affiliation{Institut Quantique, D\'epartement de physique \& RQMP, Universit\'e de Sherbrooke, Sherbrooke, Qu\'ebec, Canada J1K 2R1}
\affiliation{Canadian Institute for Advanced Research, Toronto, Ontario, Canada M5G 1Z8}

\date{\today}

\begin{abstract}

A key unexplained feature of cuprate superconductors is the upturn in their normal state electrical resistivity $\rho(T)$ seen at low temperature inside the pseudogap phase. 
We examined this issue via measurements of the thermal conductivity $\kappa(T)$ down to 50~mK and in fields up to 17~T on the cuprate La$_{2-x}$Sr$_x$CuO$_4$ at dopings $p = 0.13$, 0.136, 0.143 and 0.18.
At $p$ = 0.136, 0.143, and 0.18, we observe an initial increase of the electronic thermal conductivity \kzero~as a function of field, as expected in a $d$-wave superconductor.
For $p$ = 0.136 and 0.143, further increasing the field then leads to a decrease of \kzero, which correlates with the onset of spin density-wave order as observed in neutron scattering experiments on the same samples.
This decrease of \kzero~with field is imposed by the Wiedemann-Franz law and the high value of the resistivity in the high-field normal state of these samples.
Our study therefore provides a direct link between magnetism and the resistivity upturn in the pseudogap phase of cuprates.
We discuss this scenario in the broader context of other cuprates.

\end{abstract}


\maketitle


Across the doping phase diagram of cuprate superconductors, the normal-state resistivity $\rho(T)$ evolves through four different regimes.
At high doping, beyond the superconductivity dome (Fig.~1(a)), $\rho(T) \sim T^2$, the expected $T$ dependence of a Fermi liquid, 
as seen in La$_{2-x}$Sr$_x$CuO$_4$ (LSCO) at $p = 0.33$~\cite{nakamae2003}
and Tl$_2$Ba$_2$CuO$_{6 + \delta}$ (Tl2201) at $p \simeq 0.32$~\cite{manako1992}. 
At lower doping, just above the critical doping \pstar~for the onset of the pseudogap phase (Fig.~1(a)), 
$\rho(T)$ exhibits a $T$-linear dependence as $T \to 0$, 
as seen in 
LSCO~\cite{boebinger1996,cooper2009},
La$_{1.6-x}$Nd$_{0.4}$Sr$_x$CuO$_4$ (Nd-LSCO)~\cite{daou2009,collignon2017},
and
Bi$_2$Sr$_2$CaCu$_2$O$_{8 + \delta}$ (Bi2212)~\cite{legros2019}.
Just below \pstar~is a third regime where $\rho(T)$ is $T$-linear at high $T$ but deviates from this linear dependence below a temperature \Trho.
In LSCO and Nd-LSCO, the deviation is upwards, and \Trho~=~\Tstar~\cite{cyr-choiniere2018}, the temperature at which the pseudogap is seen to open in angle-resolved photoemission spectroscopy measurements~\cite{yoshida2009,matt2015}.

In this third regime, $\rho(T)$ shows an upturn at low temperature in some doping range, and this upturn ends by a saturation at $T \to 0$~\cite{laliberte2016}. 
In other words, the ground state is not an insulator but a metal.
Localization ($k_{\mathrm F} l \sim 1$), as evidenced by a log(1/$T$) divergence of the resistivity at low temperatures, occurs in a fourth regime, for $p \ll p^{\star}$, as the system approaches the Mott insulator at $p = 0$.

In this Letter we focus on the third regime, bounded between $p \sim 0.08$ and \pstar.
A natural mechanism for the upturn in $\rho(T)$ has recently been identified~\cite{collignon2017,michon2018,laliberte2016}: 
 there is a loss of carrier density as doping is lowered below \pstar, as inferred from the Hall number in YBa$_2$Cu$_3$O$_y$ (YBCO)~\cite{badoux2016} and Nd-LSCO~\cite{collignon2017},
which drops rapidly below \pstar~from \nH~$\simeq 1 + p$ to \nH~$\simeq p$.
In Nd-LSCO, this metal-to-metal transition with decreasing $p$ also shows up as a large and sudden drop at \pstar~in the $T=0$ electronic thermal conductivity~\cite{michon2018}.

The puzzle is this: while the low-$T$ upturn in $\rho(T)$ appears immediately below \pstar~$=0.23$ in Nd-LSCO~\cite{collignon2017}
and \pstar~$\simeq 0.18$ in LSCO~\cite{boebinger1996,momono1994}, in YBCO it appears only far below \pstar~$= 0.19$, namely below $p \simeq 0.08$~\cite{leboeuf2011}.
This means that the pseudogap phase by itself is not sufficient to cause the upturns,
something else is needed.
Here we show that magnetism is this additional necessary ingredient.
We base this conclusion on measurements of the $T = 0$ electronic thermal conductivity in LSCO that, as a function of field, exhibits a decrease above a threshold field at which prior neutron scattering studies~\cite{khaykovich2005,chang2008} observe the onset of spin density-wave (SDW) order. Taken as a proxy for the normal-state electrical resistivity, this thermal conductivity decrease corresponds to an upturn in resistivity, showing that it occurs with the onset of SDW order.

%
\begin{figure}[t!]
\includegraphics[width=0.49\textwidth]{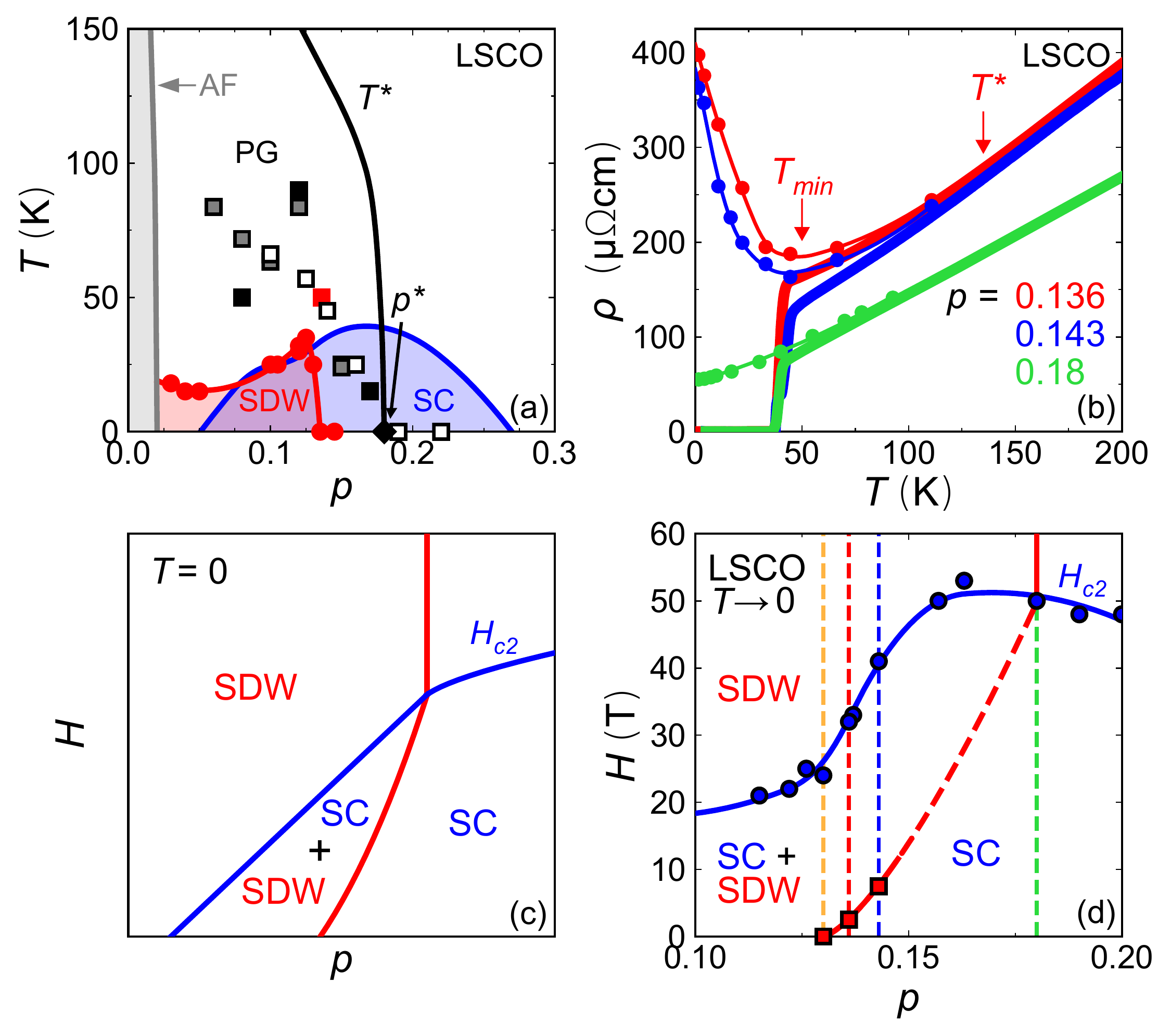}
\caption{
(a)
Temperature-doping phase diagram of LSCO, showing the Mott antiferromagnetic phase at low doping (AF), 
the superconducting phase (SC), and the pseudogap temperature \Tstar~ending at the critical doping \pstar~$\simeq 0.18$ (diamond). 
The red dots are neutron data from~\cite{kofu2009} (and references therein) that delineate the SDW phase in zero field ($H = 0$). 
Squares mark \Tmin, the temperature at which the upturn in $\rho(T)$ begins, when superconductivity is suppressed by field
(black squares~\cite{boebinger1996}; red square, panel b), Zn impurities (white squares~\cite{momono1994}), or both
(grey squares~\cite{komiya2004}).
(b) 
Electrical resistivity $\rho(T)$ of three of our LSCO samples, with dopings as indicated (from~\cite{laliberte2016}), in zero field (thick lines) and $H = 55$~T (dots). 
(For $p = 0.18$, $H = 66$~T.) 
The arrows mark \Tmin~and \Tstar~for $p = 0.136$.
(c) 
Theoretical magnetic field-doping phase diagram at $T=0$, illustrating the phase competition between SDW order and superconductivity (SC)~\cite{demler2001}.
The coexistence region (SDW+SC) is bounded on the right by a line that bends back (to the left) as the field is reduced below \Hc. 
(d)  
Experimental magnetic field-doping phase diagram for LSCO at $T \to 0$, based on our \Hc~data (blue circles) and prior neutron data (red squares~\cite{khaykovich2005,chang2008}). 
The vertical dashed lines indicate the dopings of the four samples in this study.
}
\label{Fig1}
\end{figure}
%


{\it Methods.} -- 
Large single crystals of LSCO were grown by the flux-zone technique, with nominal Sr
concentrations of 
$x = 0.13$ at the University of Bristol, 
$x = 0.144$ at Tohoku University, 
$x = 0.145$ at Hokkaido University,
and 
$x = 0.18$ at the University of Tokyo.
Neutron scattering measurements were previously performed on the large crystals with $x = 0.144$ and $x = 0.145$, 
as reported in~\cite{khaykovich2005} and~\cite{chang2008}, respectively.
For our transport measurements, a small sample was cut from each of these large crystals. 
The three samples with $x = 0.144$, 0.145 and 0.18 are the same as those used 
in our previous study of the electrical resistivity~\cite{laliberte2016}.

The long rods that are produced by the flux-zone technique typically have a variation of Sr concentration along
their length, so that the hole concentration (doping) at a particular point may not
be equal to the nominal (average) value of $x$.
The hole doping $p$ was determined as follows.
For the sample with $x = 0.18$ we take $p = x$, the Sr content.
For the samples with $x$ = 0.13, 0.144, and 0.145, where the value of $p$ needs to be known precisely,
we use the doping dependence of the tetragonal-to-orthorhombic structural transition temperature, $T_{\rm LTO}$.
The signature of $T_{\rm LTO}$ in the resistivity is a small but sharp kink~\cite{ando2004}.
The values measured in our samples are $T_{\rm LTO}$ = 228, 214
and 197~K for $x$ = 0.13, 0.144 and 0.145, respectively.
To convert into $p$, we use the linear doping dependence of the anomaly reported in~\cite{ando2004},
in the range $0.10 < x < 0.17$.
We obtain $p = 0.13$ ($x = 0.13$), 0.136 ($x = 0.144$) and 0.143 ($x = 0.145$).
The uncertainty on the value of $p$ comes from the uncertainty in determining
$T_{\rm LTO}$ from the kink in $\rho(T)$, which we estimate to be $\pm~5$~K.
This translates into an uncertainty on $p$ of $\pm~0.002$.

The \Tc~value of our samples, defined as the temperature below which the zero-field resistance is zero, is 
\Tc~$= 32.3$, 36.0, 37.3, and 35.4~K for $p = 0.13$, 0.136, 0.143, and 0.18, respectively.
%
The upper critical field \Hc~of LSCO was estimated by extrapolating \Hvs$(T)$ to $T=0$ for several LSCO samples with dopings 
in the range $0.10 < p < 0.20$ (Fig.~1(d)), where \Hvs$(T)$~is the field below which the resistivity is zero at a given temperature $T$ 
(as detailed in~\cite{grissonnanche2014}).

The thermal conductivity was measured in the field-cooled state in a dilution refrigerator over the range 50~mK to 1.0~K, 
with a one-heater-two-thermometers steady-state technique,
using the same contacts as for the electrical resistivity, 
which removes the uncertainty on the geometric factors when comparing the two.
Details of the contact preparation can be found in~\cite{laliberte2016}.
The heat current was applied in the basal plane of the orthorhombic structure of LSCO 
and the magnetic field was applied along the $c$ axis.
%


{\it Resistivity.} --
In Fig.~\ref{Fig1}(b), 
we reproduce the in-plane resistivity $\rho(T)$ of our LSCO samples at $p = 0.136$, $0.143$ and $0.18$ (from~\cite{laliberte2016}).
At high temperature, the zero-field curves all exhibit the $T$-linear behaviour characteristic of cuprates. 
At $p = 0.18$, this $T$-linear trend continues down to \Tc, while at $p = 0.136$ and 0.143 it stops at a temperature \Trho~marked by an arrow, 
which closely matches the pseudogap temperature \Tstar~\cite{cyr-choiniere2018}.
In a field large enough to fully suppress superconductivity ($H = 55$~T for $p = 0.136$ and 0.143; $H = 66$~T for $p = 0.18$), 
the normal-state resistivity exhibits a pronounced upturn at low temperature (below $\sim 50$~K) for $p = 0.136$ and 0.143, 
whereas it continues to decrease monotonically at $p = 0.18$, in agreement with prior data at similar dopings~\cite{boebinger1996,cooper2009}.
As discussed in~\cite{laliberte2016}, 
the upturns at $p = 0.136$ and 0.143 are not insulating in the sense that they saturate at a finite value as $T \to 0$,
which we label $\rho(0)$. 
In $p = 0.136$ and 0.143 at $H = 55$~T, 
$\rho(0) = 412$ and 385 $\mu\Omega$~cm, respectively. 
By contrast, at $p = 0.18$ and $H = 66$~T, $\rho(0) = 58~\mu\Omega$~cm, 
which is about 7 times smaller than the value at $p = 0.136$.


\begin{figure}[t]
\includegraphics[width=0.5\textwidth]{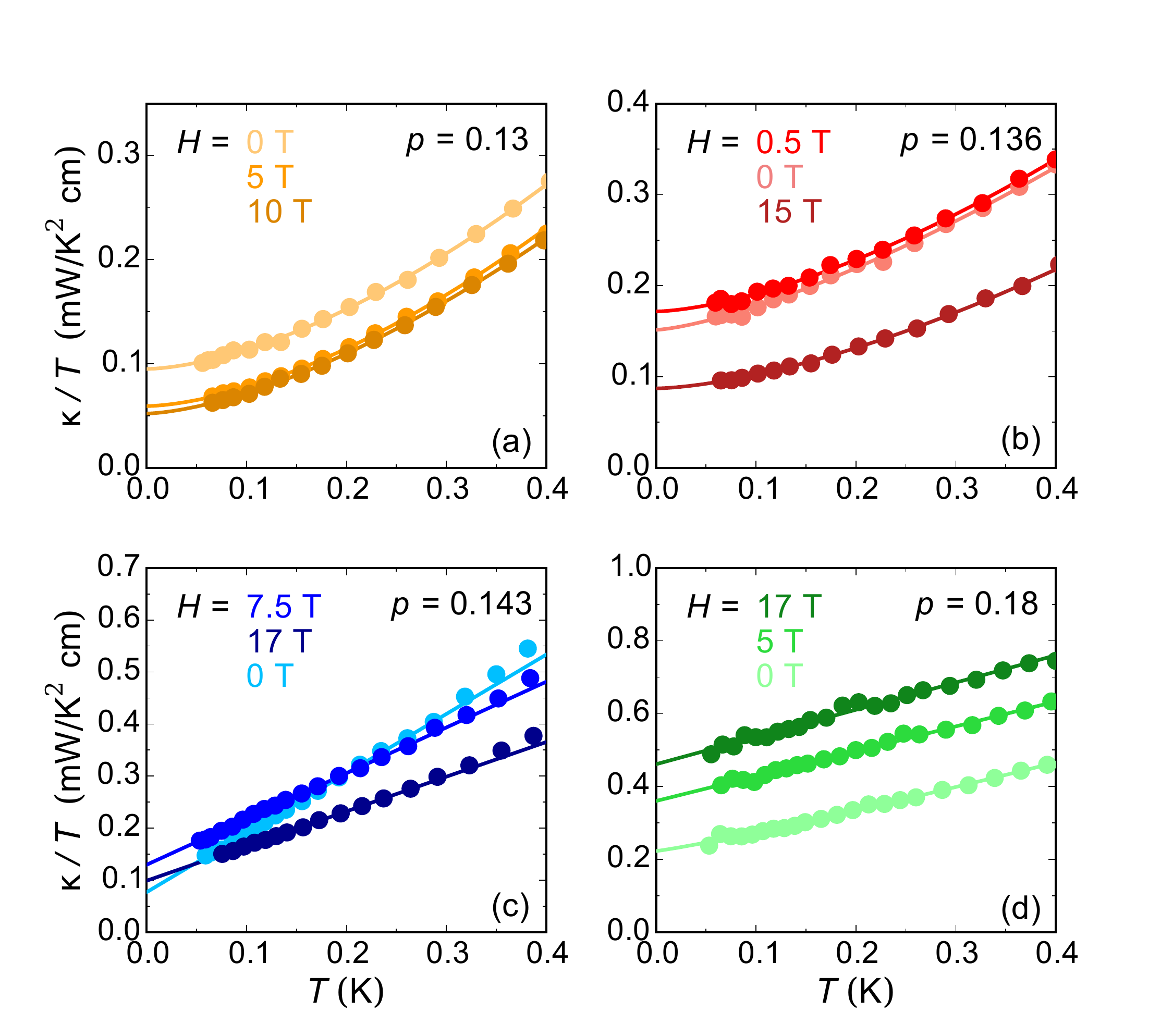}
\caption{
Thermal conductivity $\kappa(T)$ plotted as $\kappa/T$ vs $T$, at dopings and fields as indicated. 
In all panels, the lines are power law fits of the form $\kappa/T = \kappa_0/T + BT^{\alpha}$ over the entire temperature range shown. 
For $p = 0.18$, $\alpha = 1$. 
For the other dopings, $1 < \alpha < 1.7$. 
Details of the fits and values for $\alpha$ are given in Supplementary Material.
}
\label{Fig2}
\end{figure}



\begin{figure}[t!]
\includegraphics[width=0.46\textwidth]{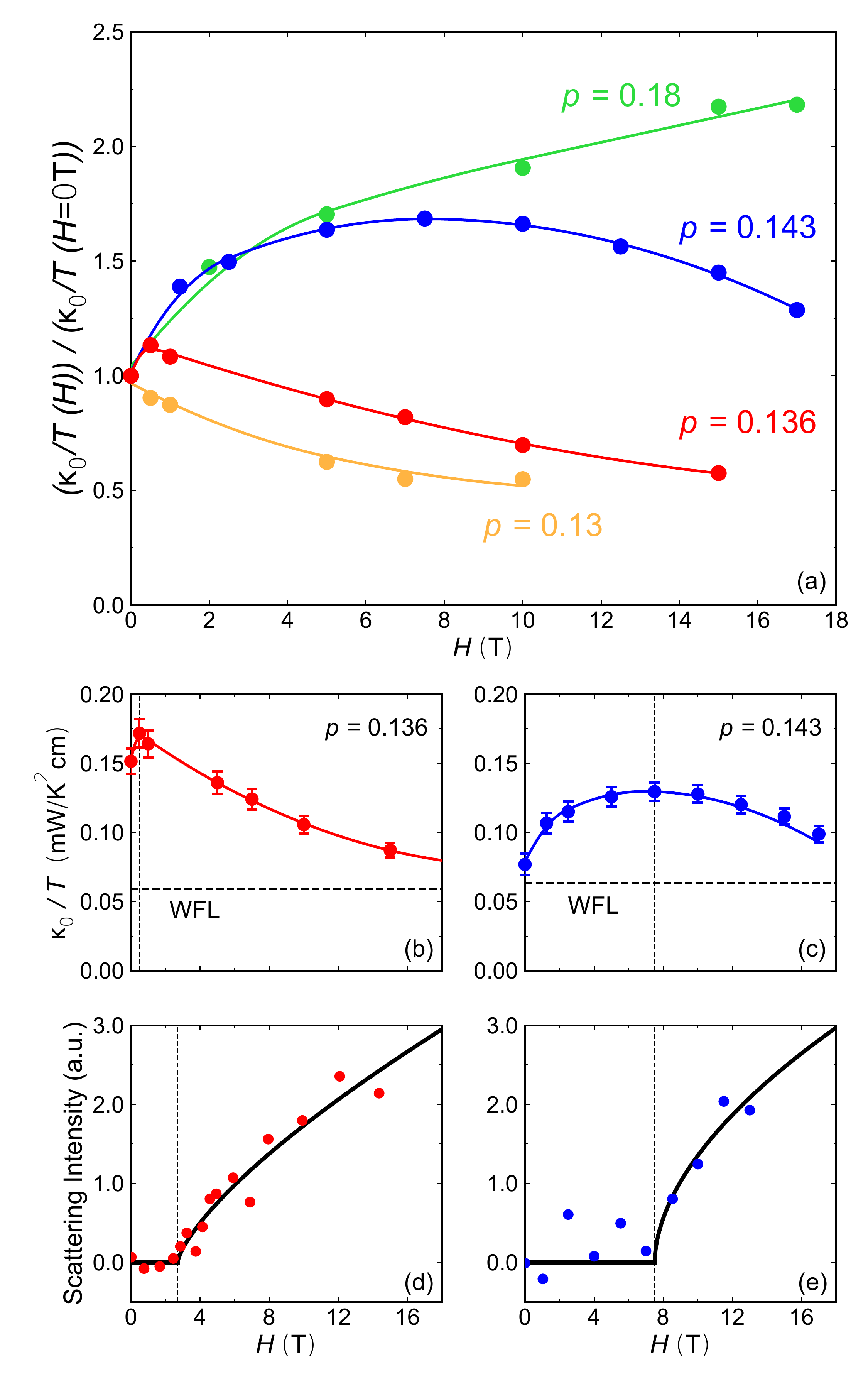}
\caption{
(a) 
Electronic thermal conductivity \kzero~as a function of magnetic field for our four samples, 
normalized to its value at $H = 0$. 
The data are extracted from the fits to $\kappa/T$ vs $T$ shown in Fig.~\ref{Fig2} and in Supplementary Material.
(b)
\kzero~vs $H$ for our sample with $p = 0.136$. 
The horizontal dashed line marks the value $L_0 / \rho(0)$ obtained from the WF law using the resistivity $\rho(0)$ at $H = 55$~T and $T \to 0$ (Fig.~\ref{Fig1}(b)). 
The vertical dashed line indicates the position of the maximum in \kzero~vs $H$. 
(c)
Same as (b), for our sample with $p = 0.143$.
(d) 
Intensity of neutron scattering at the SDW wavevector as a function of field, for the large LSCO single crystal with $x = 0.144$ (from~\cite{khaykovich2005}),
from which our sample with $p = 0.136$ was cut.
The vertical dashed line marks the threshold field above which SDW order appears, \Hsdw~$=2.5$~T.
%
(e)
Same as (d), for the large LSCO single crystal with $x = 0.145$ (from~\cite{chang2008}), 
from which our sample with $p = 0.143$ was cut.
Here, \Hsdw~$=7.5$~T.
}
\label{Fig3}
\end{figure}


{\it Thermal conductivity.} --
Our data for the thermal conductivity $\kappa(T)$ of our four LSCO samples are shown in Fig.~\ref{Fig2} and plotted as $\kappa/T$ vs $T$.
As shown by the fits, the data below 0.4~K are well described by a power-law of the form $\kappa/T = \kappa_0/T + BT^{\alpha}$, where $\kappa_0/T$ is the electronic term and $B T^{\alpha}$ is the phonon term. 
The phonon conductivity $\kappa_{\rm ph}$ goes as $\kappa_{\rm ph}/T \sim T^{\alpha}$, with $\alpha$ close to unity at high doping ($p$ = 0.18) where the system is a good metal
and phonons are mainly scattered by electrons, as in overdoped Tl2201~\cite{hawthorn2007}.
The slope $B$ is larger at $H=0$ because the density of quasiparticles that scatter phonons is lower in the superconducting state.
At low doping, where the system is much more resistive, $\alpha \simeq 1.3 - 1.7$ ($p$ = 0.13 and 0.136). 
(The values of $\alpha$~for all our samples are given in Supplementary Material.)
In Fig.~\ref{Fig3}(a), we plot the electronic term \kzero~as a function of field for our four samples, normalized by the value at $H$ = 0. 
(Data at all fields are shown in Supplementary Material.)
In all four samples, the normal state is reached only at much larger fields, between 25 and 50~T (Fig.~1(d)), 
so the data are all in the superconducting state.
For our $p$ = 0.18 sample, we observe a behavior typical of a $d$-wave superconductor, namely an increase with field over the entire range, with an initial $H^{1/2}$ dependence coming from the Volovik effect~\cite{shakeripour2009}.
The three other samples display a markedly different evolution with field.
As discovered earlier for underdoped LSCO samples~\cite{hawthorn2003,sun2003}, we observe that \kzero~decreases with $H$ at $p = 0.13$ (Fig.~3(a)).
This shows that quasiparticles in the superconducting state are more conductive than in the normal state at high field, evidence for a highly resistive normal state.

What has not been seen before is the behavior we observe at $p = 0.143$ (Fig.~3(a)): after an initial increase, \kzero~goes through a maximum at $H = 7.5$~T, and then decreases at high field.
The same is observed at $p = 0.136$, with a lower peak field, at $H \simeq 0.5-1.0$~T. 
This is our main finding.
This non-monotonic behaviour suggests the following scenario: the initial growth points to a highly metallic normal state like that seen at $p = 0.18$ (with no resistivity upturn), which gets cut-off at a certain threshold field 
by a mechanism that induces a decrease in conductivity, i.e., an upturn in resistivity.

To connect the high-field normal-state resistivity in Fig.~\ref{Fig1}(b) with our low-field thermal conductivity data, we use the Wiedemann-Franz (WF) law, given by :
\[ \frac{\kappa_0}{T} = \frac{L_0}{\rho(0)} \]
where $\rho(0)$ is the electrical resistivity as $T \to 0$ and $L_0 = (\pi^2 / 3) (k_{\rm B} / e)^2 = 2.44 \times 10^{-8}$~W~$\Omega$~K$^{-2}$.
Using the measured values $\rho(0) = 412$ and 385 $\mu \Omega$~cm at $H = 55$~T for $p = 0.136$ and 0.143, 
we obtain \kzero~= 0.059 and 0.063 mW/K$^2$~cm, respectively.
Those values of \kzero~are indicated as horizontal dashed lines in Figs.~\ref{Fig3}(b) and~\ref{Fig3}(c).
In both samples, we see that after its initial increase with field, \kzero~decreases towards a value consistent with the WF law.
This implies that whatever mechanism causes the decrease in \kzero~above a threshold field is largely responsible for the upturn in the resistivity, as suggested qualitatively in ref.~\cite{sun2003}.

{\it Spin density-wave order.} --
In LSCO at $H$ = 0, neutron scattering studies have established that SDW order exists up to a doping $p_{\rm{SDW}} \simeq 0.13$ (ref.~\cite{kofu2009} and references therein), 
as delineated by the red data points in Fig.~\ref{Fig1}(a).
However, SDW order and superconductivity compete~\cite{demler2001,katano2000,lake2002}, 
and by suppressing superconductivity with a magnetic field one can re-enter the SDW phase above $p_{\rm{SDW}} \simeq 0.13$~\cite{chang2008}.
In Figs.~\ref{Fig3}(d),(e) we reproduce neutron scattering measurements as a function of field at $p$ = 0.136 and 0.143, 
which revealed SDW order above $H_{\rm{SDW}} \simeq$ 3 and 7.5~T, respectively~\cite{khaykovich2005,chang2008}.
Our samples at $p$ = 0.136 and 0.143 were cut from the same large pieces on which these neutron scattering experiments were performed.
As shown by the vertical dashed lines in Figs.~\ref{Fig3}(b),(c), we see that $H_{\rm{SDW}}$ seen by neutrons matches well the position of the maximum in \kzero.
(We observe a slight discrepancy for the $p$ = 0.136 sample, which we attribute to a small variation in the doping of our sample with respect to the larger piece measured by neutrons, combined with the fact that the $H_{\rm{SDW}}(p)$ line in the field-doping plane has a fast slope.)
This direct correlation between $H_{\rm{SDW}}$ and the field for the peak in \kzero~is compelling evidence that the SDW order is causing the upturn in the resistivity.
Consistently, at $p$ = 0.13, SDW is present in zero field, and \kzero~always decreases with field.

Given that $p_{\rm{SDW}}$ goes from 0.13 to 0.143 in a field of 7.5~T~\cite{chang2008}, one can reasonably assume that $p_{\rm{SDW}}$ will keep increasing until $H_{c2} \sim$~50~T is reached (Fig.~\ref{Fig1}(d)).
An extrapolation of neutron data is not inconsistent with a final $p_{\rm{SDW}} \simeq 0.18$ in the complete absence of superconductivity, 
matching roughly the value for the onset of the pseudogap in LSCO, \pstar~$\simeq 0.18$.
Recent high-field nuclear magnetic resonance and ultrasound measurements show that low-$T$ magnetism does extend up to $p \simeq 0.19$ in LSCO~\cite{Frachet2019}.
For fields higher than $H_{c2}$, where resistivity measurements are performed, SDW order no longer competes with superconductivity and presumably becomes essentially field-independent.
(At $p = 0.136$, the resistivity at $T = 1.5$~K saturates above $H = 55$~T~\cite{laliberte2016}.)
This evolution as a function of doping and field for our four samples is illustrated by the sloping dashed line in Fig.~\ref{Fig1}(d).

Zn doping provides another route to supress superconductivity, and studies of Zn-doped LSCO confirm the above scenario.
Muon spin relaxation ($\mu$SR) measurements~\cite{panagopoulos2003} on pure LSCO show that spin freezing occurs at low temperatures for dopings up to $p = 0.125$, but not at $p = 0.15$, consistent with $p_{\rm{SDW}} \simeq 0.13$ as seen by neutrons.
Upon Zn-doping LSCO, neutrons scattering measurements show that a small amount of Zn induces magnetic order at $p = 0.15$~\cite{Kimura2003}, and for the highest Zn concentration $\mu$SR shows that spin freezing occurs up to $p \simeq 0.19$~\cite{panagopoulos2004}, but not above.
Electrical resistivity measurements on Zn-doped LSCO~\cite{momono1994} also show that the low-$T$ resistivity upturn occurs up to at least $p = 0.16$, but not at $p = 0.18$.
So whether superconductivity in LSCO is suppressed by a magnetic field or by Zn doping, the connection between resistivity upturn and magnetism appears to be robust, and the presence of low-$T$ magnetism throughout the pseudogap phase, up to \pstar, also.
As noted previously~\cite{komiya2004} and shown in Fig.~\ref{Fig1}(a), the onset temperature for the resistivity upturn, $T_{\rm min}$, and that for SDW order, $T_{\rm SDW}$, both peak at $p = 0.12$, further evidence of the intimate connection between magnetism and resistivity upturn.
Note that a low-$T$ resistivity upturn was observed for YBCO at $p$ = 0.18, a doping just below \pstar, in a sample whose superconductivity was weakened by electron irradiation~\cite{rullier-albenque2008}.

{\it Pseudogap phase.} --
Our inference that \psdw~=~\pstar~in LSCO also appears to be realized in Nd-LSCO,
where the onset temperature of magnetic Bragg peaks in neutron diffraction, \Tsdw, extrapolates to zero at \pstar~\cite{tranquada1997}
(as does the integrated intensity, proportional to the square of the magnetic moment).
However, note that \Tstar~$\gg$~\Tsdw, in both LSCO (Fig.~\ref{Fig1}(a)) and Nd-LSCO,
so that the pseudogap phase is not a phase of magnetic order.
We attribute the upward deviation of $\rho(T)$ below \Tstar~to the loss of carrier density in the pseudogap phase~\cite{badoux2016,collignon2017},
while the strong upturn at low $T$ requires an additional mechanism, associated with spin correlations or spin scattering.

This additional requirement is confirmed by looking at YBCO.
At $p = 0.08$, the resistivity of YBCO shows a pronounced upturn at low $T$ in $H = 55$~T, with \Tmin~$\simeq 50$~K~\cite{leboeuf2011}.
At this doping, there is SDW order at low $T$ in YBCO~\cite{haug2010}.
By contrast, in YBCO at $p = 0.11$, there is no upturn at all~\cite{leboeuf2011}, and there is also no SDW order.
(In YBCO, the presence of charge order competes with SDW order and prevents SDW order for $0.08 < p < 0.16$.)

{\it Conclusion.} --
The electronic thermal conductivity of LSCO in the $T=0$ limit is found to drop when the magnetic field exceeds \Hsdw,
the threshold field above which magnetic order appears, as seen by neutron diffraction.
This reveals a direct link between drop in conductivity -- or upturn in resistivity -- and SDW order.
We infer that magnetic order plays a key role in causing the resistivity upturns seen in the field-induced normal state of underdoped cuprates at low temperature, at least in the regime below $p^{\star}$ where the resistivity saturates at low temperature.
Extrapolation of the \Hsdw$(p)$ line up to the superconducting critical field line \Hc$(p)$ yields an intersection at $p \simeq 0.18 \simeq$~\pstar (Fig.~\ref{Fig1}(a)), the critical doping at which the pseudogap phase ends in LSCO.
In other words, \psdw~$=$~\pstar, in the absence of superconductivity at $T=0$.
This suggests that SDW order is a low-temperature instability of the pseudogap phase.

{\it Acknowledgements.} --
We thank S.~Fortier for his assistance with the experiments.
L.T. acknowledges support from the Canadian Institute for Advanced Research 
(CIFAR) as a CIFAR Fellow
and funding from 
the Institut Quantique, 
the Natural Sciences and Engineering Research Council of Canada (PIN:123817), 
the Fonds de Recherche du Qu\'ebec -- Nature et Technologies (FRQNT), 
the Canada Foundation for Innovation (CFI), 
and a Canada Research Chair.
This research was undertaken thanks in part to funding from the Canada First Research Excellence Fund
and the Gordon and Betty Moore Foundation's EPiQS Initiative (Grant GBMF5306 to L.T.).



%

\onecolumngrid

\pagebreak

\setcounter{figure}{0}
\renewcommand{\thefigure}{S\arabic{figure}}

\begin{center}
{\Large Supplementary Material for\\
``Link between magnetism and resistivity upturn in cuprates:\\
a thermal conductivity study of La$_{2-x}$Sr$_x$CuO$_4$"}
\end{center}
%
%
\begin{figure}[h]
\includegraphics[scale=1.1]{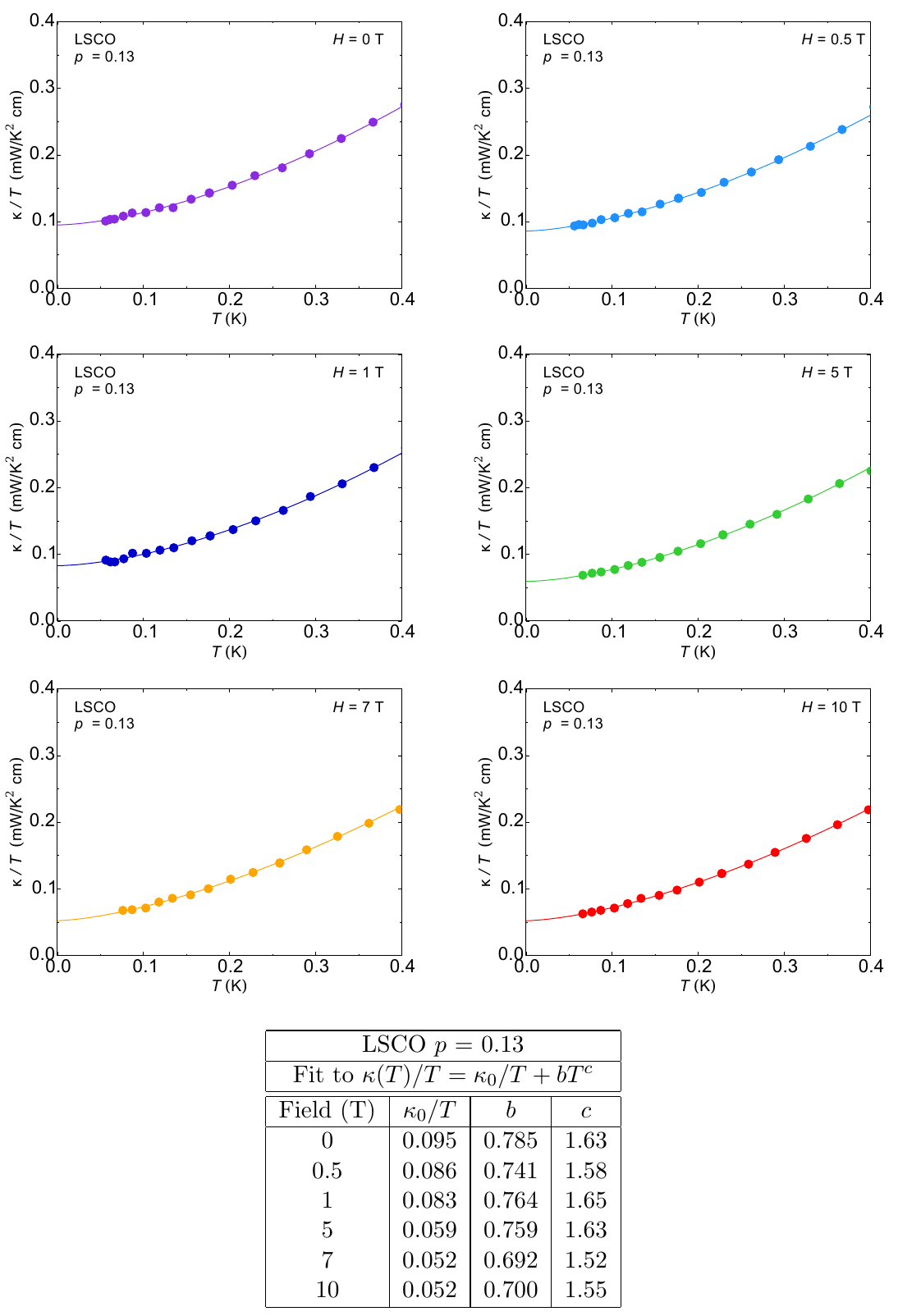}
\caption{
Thermal conductivity plotted as $\kappa/T$ vs $T$ for LSCO at $p$ = 0.13, in magnetic fields as indicated. In all panels the line is a power-law fit to the data up to 0.4~K. The fit parameters are listed in the table.
}
\label{FigS1}
\end{figure}
%
%
%
%
\begin{figure}[t]
\includegraphics[scale=1.1]{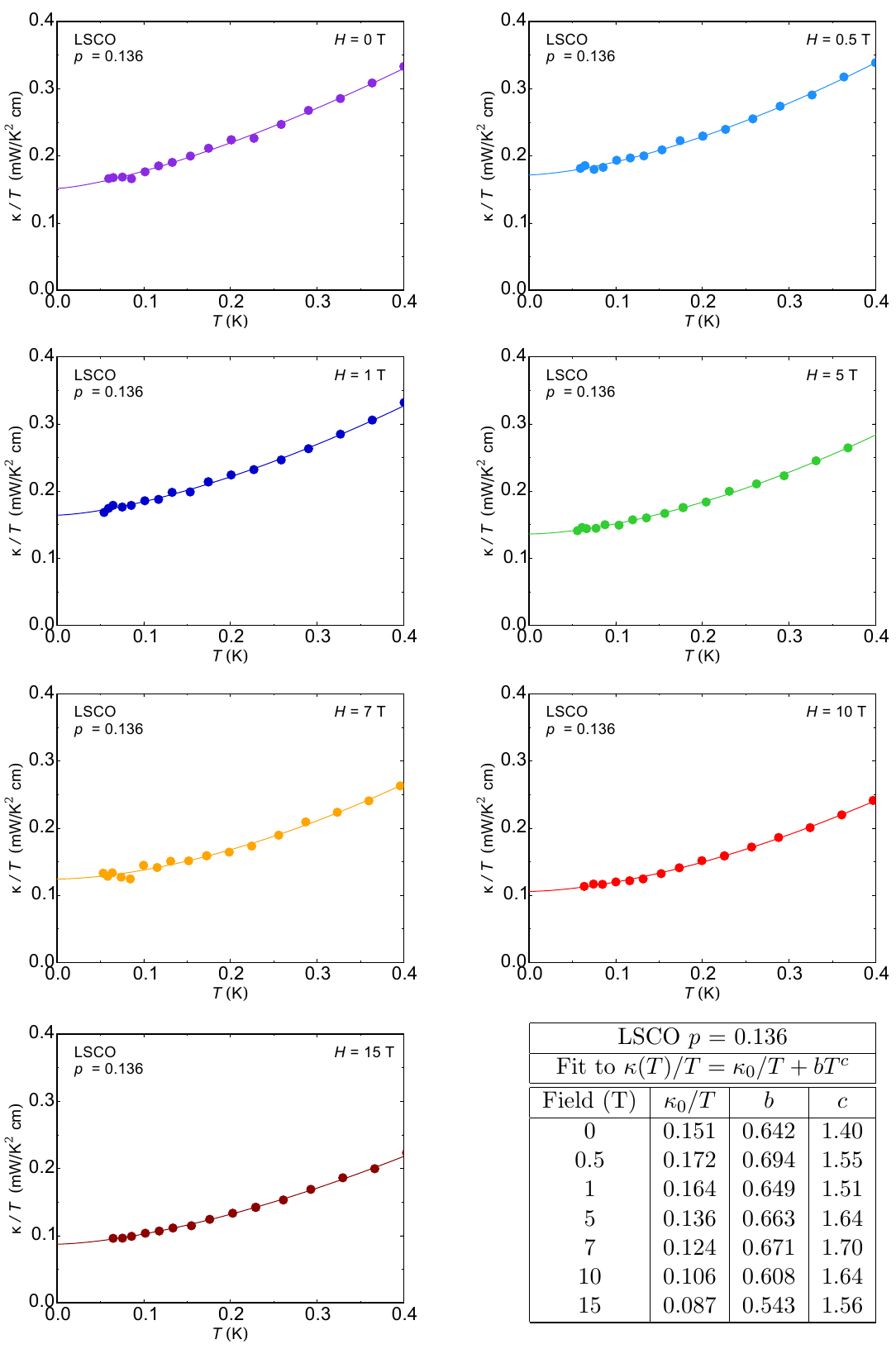}
\caption{
Thermal conductivity plotted as $\kappa/T$ vs $T$ for LSCO at $p$ = 0.136, in magnetic fields as indicated. In all panels the line is a power-law fit to the data up to 0.4~K. The fit parameters are listed in the table.
}
\label{FigS2}
\end{figure}
%
%


\begin{figure}[t]
\includegraphics[scale=1.0]{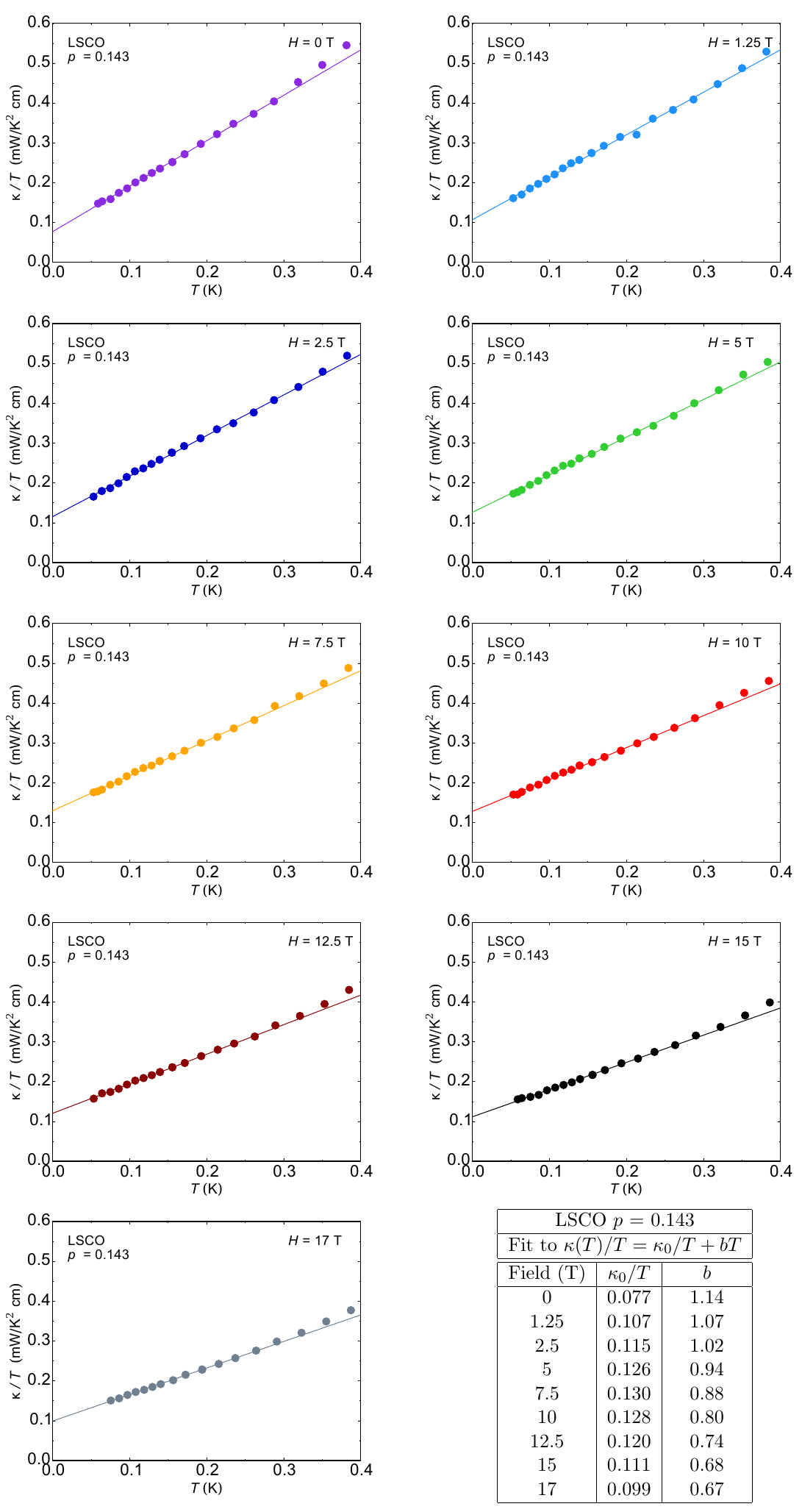}
\caption{
Thermal conductivity plotted as $\kappa/T$ vs $T$ for LSCO at $p$ = 0.143, in magnetic fields as indicated. In all panels the line is a linear fit to the data up to 0.3~K. The fit parameters are listed in the table.
}
\label{FigS3}
\end{figure}



\begin{figure}[t]
\includegraphics[scale=1.0]{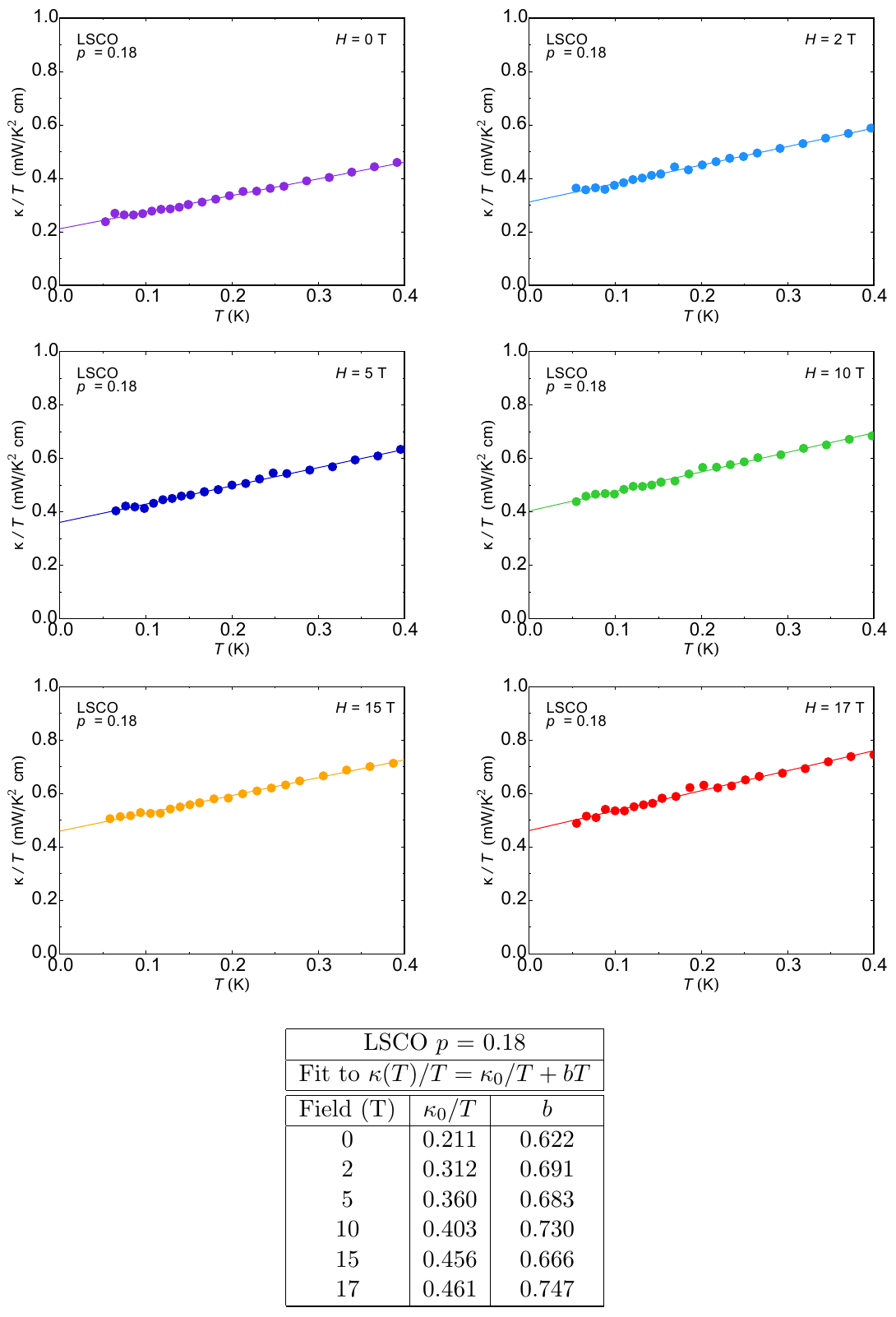}
\caption{
Thermal conductivity plotted as $\kappa/T$ vs $T$ for LSCO at $p$ = 0.18, in magnetic fields as indicated. In all panels the line is a linear fit to the data up to 0.4~K. The fit parameters are listed in the table.
}

\label{FigS4}
\end{figure}


\end{document}